\documentclass[12pt]{article}
\usepackage[cp1251]{inputenc}
\usepackage{graphicx}
\graphicspath{{pictures/}}
\DeclareGraphicsExtensions{.pdf}
\begin{document}

\title{\textbf{On Asymptotic Universality of Strong Interactions at Large Distances\footnote{Based on the talk at the 
International Conference on Quantum Field Theory, High-Energy Physics, and Cosmology.
18-21 July 2022
BLTP, JINR, Russia.To be published in PEPAN.} }}

\author{Vladimir A. Petrov
\thanks{Email:Vladimir.Petrov@ihep } \\ \\
{\it A. A. Logunov Institute for High Energy Physics} \\
{\it NRC ''Kurchatov Institute''} \\
{\it 142281 Protvino, RF } }

\date{}

\maketitle

\begin{abstract}
We propose a generalisation of the Pomeranchuk theorem which argues that elastic cross-sections should show a universal energy dependence: difference of integrated elastic cross-sections of any pair of initial channels has to disappear at high enough energy.
\end{abstract}

\section*{Terminology}

\textbf{Asymptotic} : energy $\rightarrow $ “infinity”.

\textbf{Universality}: independence of the initial collision channel.

\textbf{Strong}: inaccessible to perturbative treatment in terms of the gauge coupling.

\textbf{Distance}: average size of the interaction region ( = average impact parameter)

\textbf{Large}: noticeably larger than 1 Fermi.

\section*{Simple Example of Universality:
Gravitational Interaction}
To make things clear at once, we consider some simple and comprehensive illustration.
The gravitational potential energy of two bodies reads

\[U(r) = -G\int\frac{ d\textbf{r}'d\textbf{r}'' E_{1}(\textbf{r}')E_{2}(\textbf{r}'')}{\mid \textbf{r}-\textbf{r}'-\textbf{r}'' \mid}\]

where $E_{1,2}(\textbf{r})  $ is the energy density with
\[\int d\textbf{r}d\textbf{r} E_{1,2}(\textbf{r}) = E_{1,2} .\]

At low energies and $  r\gg R_{1,2} $ ( $ R_{1,2} $ stands for the body size)we get
\[U(r) = -G\frac{m_{1}m_{2}}{r},\]
the celebrated Newton law with an evident dependence on masses of interacting bodies.
In other extreme of high (c.m.s.) energies
$E_{1,2}\rightarrow E^{*} \gg m_{1,2}$
we get a universal expression which does not depend neither on masses nor on sizes of interacting bodies:
\[U(r)\Rightarrow -G\frac{E^{*2}}{r}\]
i.e. bears a universal character: no more dependence on specific characteristics (masses and sizes) of interacting bodies.
In a sense what will be discussed below follows \textit{grosso modo} a similar reasoning.

\section*{ Universality Types in High Energy Physics}
In 1958 Soviet physicist I.Ya.Pomeranchuk published a result\cite{bet}  which became later referred to as the "Pomeranchuk theorem" which states that 
\begin{equation}
\lim_{s\rightarrow\infty}[\sigma^{tot}_{\bar{A}B} -\sigma^{tot}_{AB}] = 0.
\end{equation}
As a theorem it assumes conditions. There are three of them:

- Analyticity in energy;

- Finite interaction radius;

- Total cross-sections tend to constant values at infinite energies.

Afterwards a weaker form of Eq.(1) which admits
cross-sections growing with energy (within the Froissart limit)was proved :
\begin{equation}
\lim_{s\rightarrow\infty}\frac{\sigma^{tot}_{\bar{A}B}}{\sigma^{tot}_{AB}}=1
\end{equation}
In 1973, Gribov predicted the universality of more general type \cite{gri}  than Eq.(2), viz.

\begin{equation}
\lim_{s\rightarrow\infty}\frac{\sigma^{tot}_{AB}}{\sigma^{tot}_{CD}}=1
\end{equation}
where $ A,B,C,D$ stand for arbitrary hadrons. 

This contrasted to then popular
statement \cite{Le} 
\begin{equation}
\lim_{s\rightarrow\infty}\frac{\sigma^{tot}_{AB}}{\sigma^{tot}_{CD}}= \frac{n_{A} n_{B} }{n_{C} n_{D} }
\end{equation}
where $ n_{A} $ is the number of valence (anti) quarks in the hadron $ A $\footnote{A transient character of Eq.(4) was clarified in \cite{pet} } .

Available experimental data  witness in favour of both Eq.(1) and, as a consequence, of Eq.(2).
As to the general conjecture (3), we seem to be at insufficiently high energy to make definite conclusions. 
Nonetheless, we believe it is of interest to consider in more detail the interpretation and underlying mechanisms of asymptotic phenomena expressed in Eqs.(1)-(3) and some further generalisations.
\section*{Some Suggestive Example and Analogy}
Hadronic interactions are known to be either elastic or inelastic.Elastic interactions (elastic scattering) are in a sense similar to collisions of billiard balls: the result of the interaction is only a deviation from the initial direction of motion of colliding objects.
From the classical viewpoint the larger the impact parameter, the weaker the interaction. For billiard balls, no interaction will occur at all if the impact parameter is greater than the diameter of the ball (we, of course, neglect their negligible gravitational attraction. On the contrary, in a head-on collision we have strong backward scattering. In principle, if the player is strong enough, then with such a frontal impact, the balls can even be destroyed (accordingly, we would deal with an inelastic event in such a case).
The picture would change if we placed electric charges on the balls. In this case we would observe deviations even at large impact parameters as electromagnetic forces have infinite range due to masslessness of the photon. Finally, if the billiard is immersed in a solution of electrolyte (or plasma), then the interaction radius becomes finite ("Debye-H\"{u}ckel radius"). Something similar we have in our case.
Strong interactions are provided by  massless gluon fields (quarkic exchanges are also possible) but due to confining structure of QCD vacuum in which interacting hadrons are immersed and  which plays the role of some specific medium, gluon fields are being screened and transformed into finite range forces. We associate them with vacuum (total singlet) Regge trajectories, the Pomerons, C-odd partners of the Pomeron, the Odderons, and "secondary trajectories" insignificant at high enough energies. As in case of electrolyte solutions (or plasma) we get finite interaction radius.
The Debye-H\"{u}ckel radius depends on the temperature. In the case of strong forces the interaction radius depends on the collision energy.

\section*{Strong Field of the Nucleon  Dependent on its Energy}

We give a simplified picture of how this happens.
At rest or at low energy we imagine the nucleon as some (fuzzy) ball formed by three valence quarks. With the growth of energy the gluon fluctuations begin to grow from inside this "valence core" and at some energy crawl out and continue to grow leaving the valence core inside.
If to take the parameters from Ref.\cite{pet}
we obtain that this happens when the nucleon energy achieves  $ 10-30 GeV $.  In terms of the c.m.s. collision energy this gives $ 20-60 GeV $ which is the ISR energy range where the elastic proton-proton cross-section starts to grow\footnote{The total cross-section starts to grow at lower energy due to the ever growing inelastic cross-section.} .
Impact parameters cannot be measured experimentally. We only can judge them with use of the differential cross-sections.
Three types of averages impact parameters are being considered which are related by one expression
\begin{equation}
\langle b^{2}\rangle_{tot} \sigma_{tot}= \langle b^{2}\rangle_{el} \sigma_{el} + \langle b^{2}\rangle_{inel} \sigma_{inel}. 
\end{equation}
We are interested in $ \langle b^{2}\rangle_{el} $ because with growing energy it grows, i.e. the short range interaction becomes more and more long range.
This quantity is defined as \cite{Ku}

\begin{center}
$ \langle b^2\rangle_{\mbox{el}} =
\bigl\langle (-t)B^2(s,t)+ 4(-t)[\partial
\Phi(s,t)/\partial t]^2\bigr\rangle _{\mbox{el}}$
\end{center}
where $ B(s,t) $ is the local slope of the differential cross-section, while $ \Phi(s,t) $ is the phase of the scattering amplitude. The averages $ \langle ...\rangle_{el} $ are provided by the probability density
\[w(t) = \frac{d\sigma_{el}}{\sigma_{el}dt}.\]

If to imagine an arbitrarily high energy we see that strong interaction becomes more and more close to the electromagnetic or the gravitational one ( distracting from the intensity of the interaction).
Therefore, it is natural to assume that the elastic cross sections depend less and less on the details of the internal structure and the general characteristics of colliding hadrons.
This conjecture is quantified as follows
\begin{equation}
\lim_{s\rightarrow\infty}[\sigma^{el}_{AB} -\sigma^{el}_{CD}] = 0,  \forall A,B,C,D.
\end{equation}
This is a generalization of the Gribov universality , Eq.(4), and is quite similar in spirit to our gravitational example above.

Why do we argue in terms of elastic cross sections, not the total or inelastic ones?
The matter is that inelastic interaction proceed more readily at shorter distances(favourable for maximum destruction of colliding hadrons)and this implies greater dependence on the type of colliding hadrons. This also concerns the total cross sections containing the inelastic ones as a significant part.

\section*{Conclusion}

Our conjecture, Eq.(6), critically depends on the character of the hadron elastic scattering. If it bears the peripheral character then our argument seems well sounded. 
In this case a non-zero difference would mean that the interacting hadrons, 
no matter how far apart they are from each other, keep information about the specific features of each other. Such a phenomenon not only looks unnatural, but also contradicts the exponential decrease at large impact parameters.

If to accept that the elastic scattering is of central character then Eq.(6) has no natural grounds and we would be forced to agree with the crazy picture of the nucleon which acquires a shape of empty  egg when achieving some energy.
This image is cherished by members of the "Holy Hollowness Sect".

Meanwhile, the juxtaposition of the "peripheral" to the "central" mode is still being discussed, although recently it has not been so active.
By now experimental data still cannot discern this dichotomy \cite{8} . 
\section*{Acknowledgement}
I am grateful to the organizers of this conference for their hospitality and well-chosen , interesting program.


\begin{thebibliography}{4}
\bibitem{bet}
\textit{I. Ya. Pomeranchuk},
Equality of total nucleon interaction cross sections
and antinucleons at high energies//
ZhETF.1958. V.34, P.725. 


\bibitem{gri}

\textit{V. N. Gribov},
Properties of Pomeranchuk poles, diffraction scattering and asymptotic equality of total cross-sections//
Sov.J.Nucl.Phys. 1973.V.17 P. 313.
 
Gribov's conjecture is discussed and commented in

\textit{V.V. Anisovich},
Gribov universality of hadron cross sections at ultrahigh energies //
Int.J.Mod.Phys. V.A 31.2016.P. 1645010.

\bibitem {Le}
\textit{E. M. Levin and L. L. Frankfurt},
The Quark hypothesis and relations between cross-sections at high-energies //
 Sov. JETP Lett. V.2.1965.P.65.

\bibitem{pet}
 
 \textit{V. A. Petrov and V.A.Okorokov},
The size seems to matter or where lies the "asymptopia"?// 
 Int.J.Mod.Phys. V.A 33.2018.P.1850077.
 
 \bibitem {Ku}
 
 \textit{V. Kundr\'{a}t and M. Lokaj\'{i}\v{c}ek}, 
 High-energy scattering amplitude of unpolarized and charged hadrons //
 Z. Phys. V.C63.1994.P. 619 .
 \bibitem{8}
 \textit{TOTEM Collaboration},
 Measurement of elastic pp scattering at $\sqrt{\hbox {s}} =8 TeV$ in the Coulomb-nuclear interference region: determination of the $\mathbf {\rho }$ - parameter and the total cross-section //
 Eur.Phys. J. V.C 76.2016. P. 661.

\end{thebibliography}
\end{document}